\documentclass[aps,prl,twocolumn,superscriptaddress,floatfix,amsmath,onlytext,fullfamily,textlf,longbibliography]{revtex4-2} 

\usepackage[normalem]{ulem}
\usepackage{bm} 
\usepackage{amssymb}
\usepackage{amsfonts} 
\usepackage{amsmath} 

\usepackage{graphicx} 

\usepackage{xcolor} 

\usepackage[pdftex,colorlinks=true,allcolors=blue,breaklinks=true]{hyperref}  

\usepackage[utf8]{inputenc} 

\definecolor{g-blue}{rgb}{0.83,0.95,1}
\definecolor{g-yellow}{rgb}{1,1,0.7}
\definecolor{g-green}{rgb}{0.9,1,0.9}
\definecolor{green}{rgb}{0,0.6,0}
\definecolor{cyan}{rgb}{0,0.7,0.7}
\definecolor{black}{rgb}{0,0,0}
\definecolor{grey}{rgb}{0.4,0.4,0.4}
\definecolor{nature-blue}{rgb}{0.0,0.200,0.500}

\def\blue#1{\textcolor{blue}{#1}}

\def \ed {\end{document}}
\def\Fbox#1{\vskip1ex\hbox to 8.5cm{\hfil\fboxsep0.3cm\fbox{%
		\parbox{8.0cm}{#1}}\hfil}\vskip1ex\noindent}  

\def\be{\begin{equation}}
\def\ee{\end{equation}}
\def\bea{\begin{eqnarray}}
\def\eea{\end{eqnarray}}
\def\bse{\begin{subequations}}
\def\ese{\end{subequations}}

\let \= \equiv  
\let\*\cdot 
\let\^\widehat 
\let\-\overline

\def\1{\bm1}

\def\<{\left\langle}    \def\>{\right\rangle}
\def\({\left(}          \def\){\right)}
\def\[ {\left[}         \def\]{\right]}

\newcommand{\Eq}[1]{Eq.\,(\ref{#1})}
\newcommand{\Eqs}[1]{Eqs.\,(\ref{#1})}
\newcommand{\Fig}[1]{Fig.\,\ref{#1}}

\newcommand{\B}[1]{{\bm{#1}}}
\newcommand{\C}[1]{{\mathcal{#1}}}    

\renewcommand{\sb}[1]{_{\text {#1}}}  
\def\Sb#1{_{\scriptscriptstyle\rm{#1}}}

\begin{document}
	
    \title{Correlation-enhanced interaction of a Bose--Einstein condensate \\ with parametric magnon pairs and virtual magnons}

	\author{Victor~S.~L'vov}
 	\email{victor.lvov@gmail.com}	
    \affiliation{Department of Complex Systems, Weizmann Institute of Science, Rehovot 76100, Israel}
    \affiliation{Department of Chemical and Biological Physics, Weizmann Institute of Science, Rehovot 76100, Israel}
	
	\author{Anna Pomyalov}
 	\email{anna.pomyalov@weizmann.ac.il}	
	\affiliation{Department of Chemical and Biological Physics, Weizmann Institute of Science, Rehovot 76100, Israel}
	 
	\author{Dmytro~A.~Bozhko}
	\email{dbozhko@uccs.edu}
	\affiliation{Department of Physics and Energy Science, University of Colorado Colorado Springs, Colorado Springs, Colorado 80918, USA}
    
    \author{Burkard~Hillebrands}
	\email{hilleb@physik.uni-kl.de}
	\affiliation{Fachbereich Physik and Landesforschungszentrum OPTIMAS, Rheinland-Pf\"alzische Technische Universit\"at Kaiserslautern-Landau, 67663 Kaiserslautern, Germany}

	\author{Alexander~A.~Serga}
	\email{serga@physik.uni-kl.de}
	\affiliation{Fachbereich Physik and Landesforschungszentrum OPTIMAS, Rheinland-Pf\"alzische Technische Universit\"at Kaiserslautern-Landau, 67663 Kaiserslautern, Germany}

\begin{abstract}  
Nonlinear interactions are crucial in science and engineering. Here, we investigate wave interactions in a highly nonlinear magnetic system driven by parametric pumping leading to Bose--Einstein condensation of spin-wave quanta---magnons. Using Brillouin light scattering spectroscopy in yttrium-iron garnet films, we found and identified a set of nonlinear processes resulting in off-resonant spin-wave excitations---virtual magnons. 
In particular, we discovered a dynamically-strong, correlation-enhanced four-wave interaction process of the magnon condensate with pairs of parametric magnons having opposite wavevectors and fully correlated phases.
\end{abstract}

\maketitle 

Nonlinear wave interactions govern the behavior of various systems in nature, including processes in the Earth's ocean and atmosphere \cite{Bartusek2022}, in stars \cite{Federrath2018}, and even the evolution of the Universe \cite{Martinelli2021}. In the field of magnetism, the use of nonlinear phenomena opens promising prospects for applications of magnonics and spintronics in neuromorphic computing, microwave data processing, and nanoscale wave logic circuits \cite{Koerner2022, Turenne2022, Gartside2022, Wang2020, Prokopenko2019}.
Spin waves, or magnons, in magnetically ordered materials are highly nonlinear compared to, for example, phonons or photons in solids. One of the best systems for studies of nonlinear spin-wave phenomena is the single-crystal ferrimagnetic yttrium iron garnet (YIG, $\rm{Y}_3\rm{Fe}_5\rm{O}_{12}$) material \cite{Cherepanov1993, Arsad2023}. A strong magnon nonlinearity, combined with a high quality factor of magnons in YIG, facilitates the creation, registration, and study of various interaction processes.

The phenomenon of Bose--Einstein condensation of magnons at the bottom of their frequency spectrum, observed in YIG films \cite{Demokritov2006, Safranski2017, Schneider2020, Schneider2021, Divinskiy2021, Dzyapko2016, Noack2021}, also develops through the nonlinear scattering of magnons.
In our experiments, this condensation was achieved by the parametric pumping of magnons with microwave radiation \cite{Demokritov2006, Serga2014, Clausen2015, Kreil2018, Dzyapko2016, Noack2021}. 
 
Here, we reveal several nonlinear processes that involve not only ``real'' quasiparticles---the eigenmodes of the medium---but produce, as a final result, virtual quasiparticles---out-of-resonance waves---caused by various types of nonlinear interactions. The most nontrivial process involves a pair of parametrically excited magnons and a Bose--Einstein condensate (BEC). This process is enhanced by full-phase correlations in parametric magnon pairs with opposite wavevectors $\B q$ and $-\B q$. 

Previously, virtual quasiparticles (e.g., virtual magnons) appeared in the theory of nonlinear waves \cite{Zakharov1992, Lvov1993, Nazarenko2011} only as mathematical abstractions, mediating interactions between ``real'' quasiparticles.
At the same time, the experimental observation of virtual quasiparticles is crucial for the identification of possible physical processes in nonlinear wave systems. This is especially critical when, for various reasons, the experimental methods do not allow one to observe the real quasiparticles involved in these processes. 

In our studies, we used Brillouin light scattering (BLS) spectroscopy \cite{Hillebrands2000, Serga2012} employing a triple-pass tandem Fabry-P\'{e}rot interferometer \cite{Mock1987}. Its high sensitivity enables the detection of a tiny number of magnons even at thermal noise level, and the frequency resolution (as low as 50\,MHz) is quite sufficient for experiments with magnons in the 3--15\,GHz range, which was explored in our investigation. 
The $532\,\rm{nm}$ laser beam is focused into a 20\,{\textmu m} diameter probing spot in the parametrically pumped area of the YIG film. 
By setting the incidence angle $\Theta_{q \parallel}$ of this beam \cite{Bozhko2020a}, the wavevectors of magnons propagating in the YIG film plane along the line of its intersection with the beam incidence plane are selected.
Since, in our case, the incident plane is oriented along the direction of the bias magnetic field $\B H$, magnons propagating in the same direction with wavevectors $\B q_\| \parallel \B H$ are detected. 
The change in the incidence angle from $0$ to the maximal value of $58^\circ$ in our experiment corresponds to the change in the detectable wavevectors from $0$ to $2\cdot 10^5 \, \rm{cm}^{-1}$. The wavevector resolution in our experiments is on the order of $\pm10^3\,\rm{cm}^{-1}$ \cite{Sandweg2010}. 

Microwave pumping at $13.2\,\rm{GHz}$ is applied to the in-plane magnetized 6.7\,{\textmu m}-thick YIG film using a microstrip resonator of 50\,{\textmu m} width. In order to achieve high magnon occupation numbers $n(\omega, \B q)$, the pumping is supplied in pulses of up to $20\,\rm{W}$ and duration of 1\,{\textmu s}. A sufficiently low repetition rate of $1\,\rm{kHz}$ maintains a thermal quasi-equilibrium from pulse to pulse. 

\begin{figure}[!t]
\includegraphics[width=1\columnwidth]{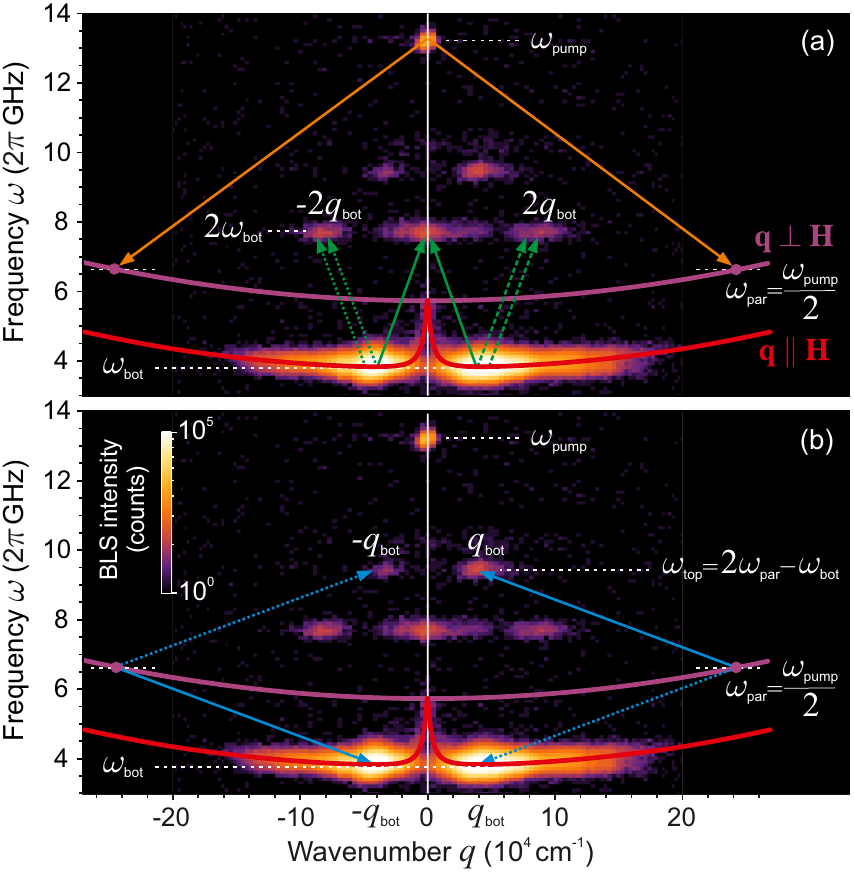}  	 
   \caption{\label{f:Experimental_spectra}Frequency- and wavevector-resolved BLS intensity spectra of real and virtual magnons. 
   The spectra were measured during the action of microwave pumping in a YIG film magnetized in-plane by the field $H=1350$\,Oe. The experimental intensity spectra are shown together with the calculated magnon dispersion curves and diagrams representing the relevant quasiparticle scattering processes in the system. (a) The process of parametric pumping is shown by orange arrows. The signal of virtual ``pump'' magnons is visible at the pumping frequency $\omega\sb{pump}$. The parametrically excited real magnons, marked by magenta dots, are outside the observation region. Virtual ``double-bottom'' magnons at the frequency $2\omega\sb{bot}$ arise due to the confluence of bottom magnons at $\omega\sb{bot}$. The three types of confluence processes are indicated by pairs of green arrows.
   b) Processes of four-wave interaction of bottom magnons with phase-correlated pairs of parametric magnons leading to the appearance of virtual ``top'' magnons at the frequency $\omega\sb{top}$ are pairwise shown by the solid and dotted blue arrows \cite{Comment_1}. 
   }
\end{figure}   
Our experimental results for the BLS intensity spectra $I(\omega, q_\|)$, proportional to  $n(\omega, q_\|)$, are shown in \Fig{f:Experimental_spectra}. 
The solid red line shows the calculated magnon frequency spectrum $\omega_{ q_\|}$ which has two minima $\omega\sb{bot}\=\min_{q_\|}\{\omega_{q_\|}\}\approx 4\cdot (2\pi)$\,GHz  at $q_{\|}=\pm q\sb{bot}$ with $q\sb{bot}\approx 4\cdot 10^3\,$cm$^{-1}$. The two brightest spots in the vicinity of the bottom of the magnon spectra at $\pm q\sb{bot}$ originate from ``bottom" magnons associated with the left and right BEC states. 

Above these brightest spots we see three spots with $\omega\approx 2\omega\sb{bot}$ and $q\sb{left}\approx -2 q\sb{bot}$, $q\sb{center}\approx 0$, and $q\sb{right}\approx 2 q\sb{bot}$. It is natural to  relate them to the confluence of two bottom magnons, as shown by green arrows in \Fig{f:Experimental_spectra}(a):  

i) left spot, $ \omega_{-q_{\|}}+ \omega_{-q_{\|}} \Rightarrow 2 \omega\sb{bot}$ and $q=-2 q\sb{bot}$; 
 
ii) central spot, $\omega_{-q_{\|}}+ \omega_{+q_{\|}} \Rightarrow 2 \omega\sb{bot}$ and $q=0$;   

iii) right spot, $\omega_{+q_{\|}}+ \omega_{+q_{\|}} \Rightarrow 2\omega\sb{bot}$ and $q= 2 q\sb{bot}$.\\
However, in our magnetization geometry, neither $\omega=2\omega\sb{bot}$ with $q=\pm 2 q\sb{bot}$ nor $\omega=2\omega\sb {bot}$ with $q=0$ are eigenmodes of the YIG film chosen. The only plausible explanation builds on the fact that we observe off-resonant waves driven by appropriate nonlinearity, i.e., virtual magnons. For brevity, we will call them ``double-bottom virtual magnons'' \cite{[{The BLS spectral peaks caused by off-resonant confluence processes can also be seen in Fig.\,3 in }]Geilen2022}.

 
In \Fig{f:Experimental_spectra}(a) the BLS spectra $I(\omega, q_\|)$ are supplemented by two down-pointing orange arrows showing the process of parametric pumping by an external quasi-homogeneous microwave field with wavevector $\B q\sb {pump}\approx 0$ and frequency $\omega\sb{pump}$. Precisely at this position, we see a rather bright spot, indicating the presence of precession with this wavevector and frequency. Since there is no corresponding magnon eigenmode with $\omega=\omega\sb{pump}$, we observe here one more type of virtual magnon, called, for concreteness, virtual ``pump'' magnons. These magnons are directly driven by the external microwave magnetic field. 

At the same time, we see no BLS response at the frequency of the parametrically pumped real magnons $\omega\sb{par}=\omega\sb{pump}/2$ because the wavenumbers of the parametric magnons are quite large and lie outside the sensitivity limit of our BLS setup at $\approx 2\cdot 10^5 \, \rm{cm}^{-1}$. Nevertheless, these magnons are definitely present in the system \cite{Serga2012}, being responsible for the appearance of magnons at the bottom of the frequency spectrum \cite{Bozhko2015, Kreil2018}, which are clearly visible as extremely intense BLS signals.  

In our previous studies \cite{Bozhko2017, Bozhko2020, Frey2021, Hahn2022}, the magnon gas thermalization and BEC formation were developing in close interactions with the phonon bath \cite{Hick2010, Ruckriegel2014, Bozhko2020}, and leading to accumulation of hybrid magnon-phonon quasiparticles. In the present experiments, we observe no such phenomenon because the width of the pumping area is small, and the strong efflux of these quasiparticles leads to large losses for them \cite{Bozhko2017}.

Two more spots at $\omega\sb{top}=\omega\sb{pump}-\omega\sb{bot}$ and $q\sb{top}=\pm q\sb{bot}$, have a more sophisticated origin, which we discuss below within the classical Hamiltonian formalism convenient for description of nonlinear processes at large occupation numbers $n(\B q)\gg 1$.

The classical Hamiltonian formalism is based on the equation of motion for the complex canonical wave amplitudes $a_{\B q }$ and $a_{ \B q}^*$, which are classical limits of the Bose-operators $ \widehat a_{\B q}$ and $\widehat a_{\B q}^\dag$ \cite{Zakharov1992, Lvov1993, Nazarenko2011}:
  \begin{equation}\label{ME}
   \frac{d a_{\B q }( t)}{d t}=-i \frac{\delta \C H}{\delta a^*_{\B q}( t)} ,
\end{equation} 
where $\C H$ is the Hamiltonian function, called for brevity the Hamiltonian. 

For weakly interacting waves, $\C H$ can be expanded in $a_{\B q }$ and $a_{\B q}^*$ as
$\C H  = \C H_2+\C H\sb{int}$. 
The quadratic Hamiltonian $\C H_2 =\sum_{\B q}\omega_{\B q}a_{\B q } a_{\B q}^*$ describes the free propagation of non-interacting waves with the dispersion law $\omega_{\B q}$: $a_{\B q}(t)\propto \exp [-i \omega_{\B q}t]$. 
The interaction Hamiltonian $\C H\sb{int}=\C H_3+\C H_4$, includes  
\begin{subequations}\label{Ham}  \begin{align}
   \label{HamB} 
\C H_3&=\frac 12 \sum_{\B 1,\B 2,\B 3}\Big [ V_{\B 1,\B 2}^{\B  3}  a_{\B 1}a_{\B 2} a_{\B 3} ^*+(V_{\B 1,\B 2}^{\B 3})^* a_{\B 1}^*a_{\B 2}^*a_{\B 3}\Big]\Delta _{\B 1,\B 2}^{\B  3}\,,\\ \label{HamC} 
\C H_4&=\frac 14 \sum_{\B 1,\B 2,\B 3,\B 4}  T_{\B 1,\B 2}^{\B 3,\B 4}  a_{\B 1}^*a_{\B 2}^*a_{\B 3}a_{\B 4} \Delta _{\B 1,\B 2}^{\B 3,\B 4} \,.
 \end{align}\end{subequations}
Here, $V_{\B 1,\B 2}^{\B  3}$ and $T_{\B 1,\B 2}^{\B 3,\B 4} = \big(T^{\B 1,\B 2}_{\B 3,\B 4}\big)^*$ are three- and four-wave interaction amplitudes. The Kronecker symbols $\Delta _{\B 1,\B 2}^{\B 3 }$ and $\Delta _{\B 1,\B 2}^{\B 3,\B 4}$ are equal to one if the sum of ``super-script'' wavevectors equals the sum of ``sub-script'' wavevectors, and zero otherwise, taking care of wavevector conservation. For shortness, hereafter we use the notation $\B j\=\B q_{\B j}$, with $j=1,\, 2,\, 3,\, 4$.
 
\begin{figure}[t]
 \includegraphics[width=1\columnwidth]{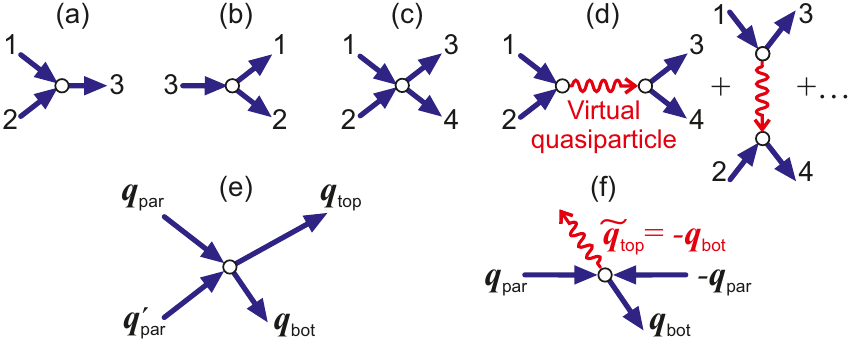}
   \caption{\label{f:Scattering_diagrams} Graphical representation of multi-wave scattering processes. Three-wave confluence (a) and splitting (b) processes given by \Eq{3}. (c)\,The four-wave scattering process given by \Eq{4w}. (d) The first (see \Eq{VM6}) and second contribution to the 4-magnon scattering process in the second-order perturbation approach in the three-wave Hamiltonian, \Eq{HamB}. The red wavy lines show virtual quasiparticles, mediating these processes. All six contributions can be found, for example, in Eq.(1.1.32) from Ref.\,\cite{Lvov1993}.
     (e)\,Diagram of the four-magnon scattering\,[\Eq{condA}] leading to the kinetic instability of real magnons. 
    (f)\,Diagram of the four-magnon scattering\,[\Eq{condA}] that leads to the excitation of the virtual ``top'' magnons.  
     }
\end{figure}

The close analogy between the classical Hamiltonian and a quantum mechanical description allows for a physically transparent interpretation of the interactions $\C H_3$ and $\C H_4$. Namely, $\C H_3$ describes the process of the confluence of two waves (or two quasiparticles, such as magnons, phonons, etc.) with the wavevectors $\B q_1$ and $\B q_2$ into one wave with the wavevector $\B q_3 = \B q_1+\B q_2$, $2\Rightarrow 1$ [see \Fig{f:Scattering_diagrams}(a)], and the inverse process of decay of the $\B q_3$-wave into a pair of $\B q_1$- and $\B q_2$-waves, $1\Rightarrow 2$ [see \Fig{f:Scattering_diagrams}(b)]:
    \begin{equation}\label{3}
            \omega_{\B 1}+ \omega_{\B 2}\Leftrightarrow \omega_{\B 3}\ .
    \end{equation}
Hamiltonian $\C H_4$ is responsible for the $2\Leftrightarrow2$ scattering [see \Fig{f:Scattering_diagrams}(c)] with the conservation laws 
\begin{equation}\label{4w}
        \omega_{\B 1} +\omega_{\B 2}\Leftrightarrow \omega_{\B 3}+\omega_{\B 4}\,,\quad  q_{\B 1} + q_{\B 2}=  q_{\B 3}+q_{\B 4}\,.
    \end{equation}
 
As a general rule, the $2\Leftrightarrow1$ processes are more efficient. When they are forbidden by the conservation law \eqref{3}, one has to account for weaker scattering processes  $2 \Leftrightarrow 2$, \Eq{4w}. Nevertheless, three-wave processes cannot be ignored in this case either. 
To see this, we examine equation \Eq{ME} with the Hamiltonian\,\eqref{Ham}: 
 \begin{align}  
    \begin{split} \label{MEQ}
    \Big [\frac d {d \B t}+  i \omega_{\B q}\Big ]a_{\B q}=& -\frac i2 \sum_{\B  1,\B  2 }      V_{\B 1,\B 2}^{\B  q} \Delta _{\B 1,\B 2}^{\B  q} a_{\B 1}a_{\B 2}  \\  
    + \sum_{\B  1,\B  2 }  (V_{\B q,\B 2}^{\B  1})^* \Delta _{\B q,\B 2}^{\B  1} a_{\B 1}^*a_{\B q}  
  &  -   \frac i2 \sum_{ \B  2,\B  3,\B  4}  T_{\B q,\B 2}^{\B  3,\B  4}  a_{\B 2}^*a_{\B 3}a_{\B 4} \Delta _{\B q,\B 2}^{\B  3,\B  4} \ .
    \end{split} 
\end{align} 
The first term in the right-hand side of \Eq{MEQ} is the sum of off-resonant forces
$ F \Sb V (\B r, t)\propto \exp\{i[\B  q\Sb V \cdot \B r]-\omega \Sb V t\} $
with the wavevector $\B  q\Sb V=\B q_1+\B q_2 $ and frequency $ \omega \Sb V=\omega _{\B 1}+\omega_{\B 2}$. This force off-resonantly stirs up the driven wave 
\begin{equation}\label{VM1} 
a (\B q_1+\B q_2)=   \sum_{\B  1,\B  2  }  V_{\B 1,\B 2}^{\B 1+\B 2}  a_{\B 1}a_{\B 2} \big [2\big (\omega_{\B 1}+\omega_{\B 2}- \omega_{\B 1+\B 2}\big ) \big ]^{-1} \ .
\end{equation}
Taking $a_{\B 1}$ and $a_{\B 2}$ as $a_{\pm q_{\|}}$, we find the amplitudes of the virtual double magnons discussed above.
By substituting $a(\B q_1+\B q_2)$ from \Eq{VM1} into \Eq{HamB} for the Hamiltonian $\C H_3$ we find an additional contribution to the four-wave interaction amplitude
\begin{equation} \label{VM6} 
    \delta T_{\B 1, \B 2}^{\B 3, \B 4} =  V_{\B 1, \B 2}^{\B 1+\B 2}{V _{\B 1+\B 2}^{\B 3,\B 4}}^*  \big ( \omega_{\B 1}+\omega_{\B 2}-\omega_{\B 1 + \B 2}\big )^{-1}\,,
\end{equation}
arising in the second order of the perturbation theory by the amplitude of three-wave processes $ V_{\B 1,\B 2}^{\B  3}$ through the virtual wave with the frequency $\omega\Sb V=\omega_{\B 1+\B 2}$:
$\omega_{\B 1}+\omega_{\B 2} \Rightarrow \omega_{\B 1+\B 2} \Rightarrow \omega_{\B 3}+\omega_{\B 4}$.  
This process is shown in \Fig{f:Scattering_diagrams}(d).

\begin{figure*}[t]
  \includegraphics[width=1 \linewidth]{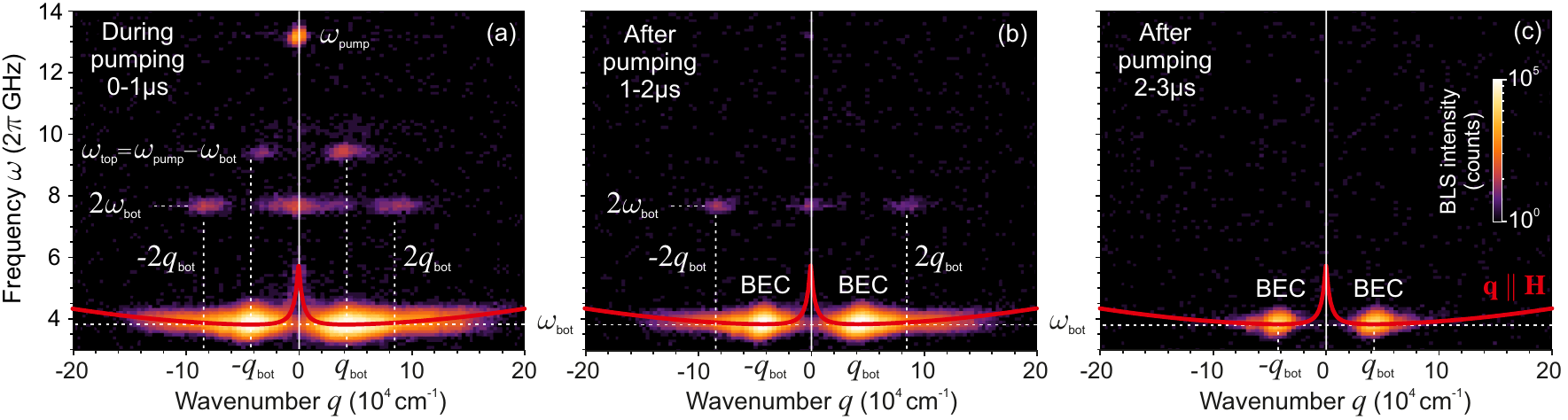}  
    \caption{\label{f:Time_evolution} Time evolution of the BLS spectra of the real and virtual magnons. (a) During the pumping microwave pulse, the virtual ``top'' magnons at $\omega\sb{top}=\omega\sb{pump}-\omega\sb{bot}$ are visible alongside with ``double-bottom'' virtual magnons at $2\omega\sb{bot}$. (b) Immediately after the pumping pulse, the virtual ``top'' magnons disappear, and only ``double-bottom'' virtual magnons are visible as the populations of magnon BECs at $\omega\sb{bot}$ are still large. (c) 2\,{\textmu s} after the pumping pulse, only magnon BECs are visible. Due to the small magnon quantity, virtual magnons are not observable. All panels: the red solid curves show the calculated dispersion curves of magnons with uniform distributions of dynamic magnetization over the thickness of the YIG film and propagating in the film plane along the magnetization field $\B H$.
	}    
\end{figure*}

One can say that the process of nonlinear interaction of $\B q_1$- and $\B q_2$-quasiparticles generates the driving force $F\Sb V (\omega,\bm{q})$ with the frequency $\omega\Sb V $ and the wavevector $\bm{q}\Sb V =\bm{q}_1+\bm{q}_2$. This force non-resonantly excites a wave perturbation in a nonlinear medium---a virtual quasiparticle with an amplitude $a_{q\Sb V}$ given by \Eq{VM1}. The virtual quasiparticle decays into two real $\B q_3$- and $\B q_4$-quasiparticles with frequencies $\omega_{\B 3}$ and $\omega_{\B 4}$ resulting in a four-wave scattering process, \Eqs{4w}. 

Returning to the analysis of our experimental results, note that in our geometry and for $H=1350\,$Oe the high population of the bottom magnons is the result of the so-called kinetic instability\,\cite{Lavrinenko1981, Melkov1991, Kreil2018}: a four-magnon process\,\eqref{4w} of decay of parametric magnons with frequency $\omega_{\B 1}=\omega_\B 2=\omega \sb{par}=\omega\sb{pump}/2$ and wavevectors $\B q_1=\B q\sb {par}$, and $\B q_2=\B q'\sb {par}$ into a pair of the bottom and the so-called ``top'' magnons with frequencies $\omega\sb {bot}$ and $\omega\sb {top}$ [see \Fig{f:Scattering_diagrams}(e)]. If so, one expects 
\begin{subequations}\label{cond}
    \begin{align}\label{condA}
       \omega_{\B 1}+ \omega_{\B 2} & =\omega\sb {bot} + \omega\sb {top} = \omega \sb{pump}\,, \ \mbox{and} \\ \label{condB}\B q_1+ \B q_2&= \pm \B q_0 + \B q\sb{top}\,.
    \end{align}
\end{subequations} 
Assuming a rough estimate that $\omega\sb{top}\gg \omega\sb{bot}$, 
we conclude that $q \sb{top} > q\sb{par}$ meaning that the top magnons lie outside the sensitivity region of our BLS setup, i.e., they are invisible in \Fig{f:Experimental_spectra}. This raises the crucial question: What is the origin of the two spots in \Fig{f:Experimental_spectra}b at the ``correct'' frequency $\omega\sb{top}=\omega\sb{pump}-\omega\sb{bot}$, which is consistent with \Eq{condA}, but with the experimentally observed wavevectors $\B q\sb{top}=\pm \B q\sb{bot}$, which completely disagree with \Eq{condB}?

To resolve this controversy, we note that the theory of kinetic instability is formulated in the framework of the weak-wave kinetic equation, which assumes weak correlations of the wave phases. 
As a result, the scattering\,\eqref{4w} of real magnons has a stochastic nature and appears only as a second-order perturbation of the four-wave interaction amplitudes $T_{\B 1,\B 2}^{\B 3, \B 4}$, \Eq{HamC}.
Nevertheless, in our particular case with a large population of parametric and bottom magnons, the scattering waves have strong externally determined phase correlations.
In particular, the full phase correlation in the pairs of parametric waves with  $\pm \B q\sb{par}$ arises due to their interaction with the space-homogeneous pumping field \cite{Lvov1993}, thus 
 $  | \< a_{\B q\sb{par}}a_{-\B q\sb{par}}
   \exp [i \omega \sb{par}] \>|= \< |a_{\B q\sb{par}}|^2 \> \=n_{\B q\sb{par}}$.
This allows us to consider a pair of parametric magnons $(a_{\B q\sb{par}}a_{-\B q\sb{par}})$ as a ``single'' wave object with the frequency $2\omega\sb{par}=\omega\sb{pump}$ and phase being the sum of the phases of the waves composing the pair. Therefore, due to its dynamic nature, the four-wave scattering process\,\eqref{4w} with $\B q_1= -\B q_2 $, $\B q_3=\pm \B q\sb{bot} $ is much stronger than stochastic scattering \eqref{4w} with $\B q_1 \ne -\B q_2$, being now proportional to the first power of the interaction amplitude $T_{\B 1,\B 2}^{\B 3, \B 4}$.
One can see this from the last term in the right-hand side of \Eq{MEQ}, which describes a driving force 
\begin{equation}
  \widetilde F\Sb V (\omega,\mp \bm{q}\sb{bot})= \sum_{ \B q\sb{par}}  T_{\mp \bm{q}\sb{bot},\pm  \bm{q}\sb{bot}}^{\B q\sb{par},\B- q\sb{par}} a_{\B 2}^*a_{\B q\sb{par}}a_{-\B q\sb{par}}  
\end{equation}
having the same frequency $\omega\sb{top} =\omega\sb{pump}-\omega\sb{bot}$ as that of real top magnons [see \Eq{condA}] and wavevector $\bm{q}\sb{top}  =\mp \B q\sb{bot}$. This force  should  off-resonantly excite  virtual magnons  with amplitude $\widetilde a\sb{top}$, given by  
\begin{equation}\label{7}
    \widetilde a\sb{top}(\omega\sb{top}, \pm\B q\sb {bot})=  \widetilde F\Sb V (\omega,\mp \bm{q}\sb{bot}) / [2(\omega\sb{bot}-\omega\sb{par})].
\end{equation}  

In \Fig{f:Experimental_spectra}(b) one can see two bright spots exactly at $\omega=\omega\sb{top}$ and $\B q \sb{top}\pm\B q\sb {bot}$.
We associate these spots with the ``top'' virtual magnons described above.  

It is worth noting that such an enhancement of the efficiency of nonlinear processes in the case of phase-correlated waves is not a specific feature of the magnon system. For example, it is known in quantum optics for simpler cases when only a few waves are involved in the interaction process \,\cite{Turschmann2019}. 

To further confirm the origin of virtual magnons, we studied the evolution of magnons after the pumping is turned off. 
Figure\,\ref{f:Time_evolution} shows the BLS spectra of the real and virtual magnons at different moments of time. Figure\,\ref{f:Time_evolution}(a) shows spectra during the pumping process (same as presented in \Fig{f:Experimental_spectra}, for reference). 
Figure\,\ref{f:Time_evolution}(b) refers to the time interval $t=(1 - 2)$\,{\textmu s} after the pumping power has been turned off and the rapidly relaxing parametric magnons have disappeared.
Therefore, we do not see the pump and top virtual magnons, forced either directly by the pumping magnetic field or through the parametric magnons, as seen in \Fig{f:Scattering_diagrams}(f). 
The signal of the top virtual magnons disappears in a time not exceeding 20\,ns. This value correlates well with earlier measurements of the scattering time of a group of parametrically pumped magnons, which in the BEC experiments was close to 10\,ns \cite{Serga2014}.
Nevertheless, the double-bottom virtual magnons are still there, forced only by the bottom real magnons. They disappear only later, see \Fig{f:Time_evolution}(c), when the number of the (real) bottom magnons becomes \blue{too} small to force the double magnons above a measurable level. 
Thus, the time evolution of the virtual magnons confirms the mechanism of their excitation suggested in our paper. 

Traditionally, nonlinear-wave interactions are investigated via observation of ``real'' quasiparticles---quanta of freely propagating waves, i.e., eigenmodes of the nonlinear medium. In our experiments with highly populated magnon gases in a ferromagnetic YIG film, we found several types of virtual magnons---quanta of off-resonant spin waves---driven by real magnons through the appropriate nonlinearity. We show that virtual quasiparticles (virtual magnons in our case) may be the end result of a new type of nonlinear processes in systems with narrow intense packets of quasiparticles. In particular, we discovered a four-wave interaction involving pairs of parametric waves with opposite wavevectors and a magnon Bose-Einstein condensate. This process is strongly enhanced by phase correlation in parametric pairs and BECs. 
The considered interactions that lead to the appearance of virtual quasiparticles  contribute to the behavior of nonlinear systems, and are useful for the identification of physical processes in them. 
Furthermore, the study of various types of virtual and real quasiparticles and their correlation-enhanced interactions in the classical and quantum limit is necessary, for example, for their application in new technologies of information transfer and data processing \cite{Fukami2021}, including coherent quantum optics, see e.g. \cite{Turschmann2019} and references therein.

This study was funded by the Deutsche Forschungsgemeinschaft (DFG, German Research Foundation) -- TRR 173/2 -- 268565370 Spin+X (Project B04). D.A.B. acknowledges support by grant ECCS-2138236 from the National Science Foundation of the United States. V.S.L. was in part supported by  NSF-BSF grant \# 2020765. We thank R.~Verba, P.~Pirro and M.~Fleischhauer, for fruitful discussions.



\bibliography{VirtualMagnons.bib} 
 
\end{document}